# The New Nitrides: Layered, Ferroelectric, Magnetic, Metallic and Superconducting Nitrides to Boost the GaN Photonics and Electronics Eco-System

Debdeep Jena[1,2]*, Ryan Page[2], Joseph Casamento[2], Phillip Dang[1], Jashan Singhal[1],

Zexuan Zhang[1], John Wright[2], Guru Khalsa[2], Yongjin Cho[1], and Huili Grace Xing[1,2]

[1]*Electrical and Computer Engineering, Cornell University, 14853, Ithaca, New York, USA*

[2]*Materials Science and Engineering, Cornell University, 14853, Ithaca, New York, USA*

[3]*Applied Physics, Cornell University, 14853, Ithaca, New York, USA*

E-mail: djena@cornell.edu

The nitride semiconductor materials GaN, AlN, and InN, and their alloys and heterostructures have been investigated extensively in the last 3 decades, leading to several technologically successful photonic and electronic devices. Just over the past few years, a number of "new" nitride materials have emerged with exciting photonic, electronic, and magnetic properties. Some examples are 2D and layered hBN and the III-V diamond analog cBN, the transition metal nitrides ScN, YN, and their alloys (e.g. ferroelectric ScAlN), piezomagnetic GaMnN, ferrimagnetic $Mn_4N$, and epitaxial superconductor/semiconductor NbN/GaN heterojunctions. This article reviews the fascinating and emerging physics and science of these new nitride materials. It also discusses their potential applications in future generations of devices that take advantage of the photonic and electronic devices eco-system based on transistors, light-emitting diodes, and lasers that have already been created by the nitride semiconductors.

## 1. Introduction

The discovery of p-type doping[1-3] of GaN with Mg around 1990 unleashed the potential of the nitride family of semiconductor materials comprising of GaN, InN, and AlN, and their alloys and heterostructures. The ability to inject holes from p-type GaN into InGaN quantum wells made blue LEDs and Lasers possible, followed by phosphor-assisted down-converted green LEDs, and white LEDs for solid state lighting. Today white LEDs are ushering in the replacement of Edison's incandescent bulbs, a long-awaited 21st century quantum-facelift commensurate with the replacement of coal and oil by solar and renewable energies. But science is an endless frontier: fundamental scientific research in nitrides is moving towards taking up the challenge of photonic light emitters at ever shorter wavelengths, from the blue (~450 nm) and visible to deep-ultraviolet, approaching 200 nm, powered by AlN of bandgap ~6.2 eV (~200 nm). The challenges are deja-vu: there are no simple known ways to p-type dope high Al-content AlGaN. This electronic conductivity problem is coupled with a photonic conductivity problem: extraction of very short wavelength light trapped inside the semiconductor material in a deep-UV LED presents a twin challenge[4] similar to what kept nitride materials and visible photonics researchers busy for the last three decades.

Close on the heels of the developments in visible photonics in the 1990s was the discovery of a 2-dimensional electron gas (2DEG) at nominally undoped AlGaN/GaN heterojunctions[5]. The formation of these 2DEGs was discovered to be due to the combination of the internal spontaneous and piezoelectric polarization fields and band offsets[6-9]. Today, the Al(Ga)N/GaN heterojunction system, with a polarization-induced 2DEG and large bandgap channel and barrier, powers nitride high-electron mobility transistors (HEMTs)[10]. The high electron mobility and saturation velocity of electrons in the 2DEG with the large breakdown voltage allows high frequency radio-frequency amplifiers, and fast high-voltage switching. Nitride HEMTs are making an impact in the arenas of microwave electronics for communications, and in energy-efficient power electronics. Recently, the discovery of undoped polarization-induced 2D hole gases in nitride heterostructures[11-12] has raised the prospects for nitride high-voltage complementary electronic logic devices[13]. The digital CMOS challenge, frontiers of power electronics, and generating larger powers at microwave and THz frequencies for future communication systems are expected to keep Nitride electronics researchers busy for the next decades.

Advances in the growth and doping of GaN, AlN, and InN based materials and their heterostructures, the exploration of the effects of spontaneous and piezoelectric polarization in these semiconductors, and advances in their processing and fabrication have laid a solid foundation for the photonic and electronic devices eco-system that the field enjoys today. But no field should rest on its laurels. As the canonical nitride semiconductor materials mature, there are a number of new nitride materials that are showing glimmers of promise to significantly expand the device application space. This article provides a brief introduction to these *new* nitride materials, their fascinating properties, and a discussion of their potential applications. The materials discussed in this article, by no means comprehensive, may be new to the GaN semiconductor community, but they have been investigated by related research communities for quite some time. Fig. 1 shows the elements of the periodic table that these new nitride materials contain, together with the traditional III-Nitrides (AlN, GaN, InN) and donors (Si, Ge) and the only known shallow acceptor (Mg). Fig. 2 shows the new nitride materials in the traditional bandgap-lattice constant format, and Fig 3 shows the various materials families for the new nitrides.

## 2. Boron-based Nitrides

Boron is the lightest group-III element. Before discussing boron nitride, an analogy to a closely related material can guide chemical and physical expectations. Between Boron and Nitrogen in the periodic table is Carbon. The stable crystal structure of carbon under atmospheric conditions at room temperature is graphite, the layered form of $sp^2$ carbon, the monolayer version of which is graphene. Diamond, the $sp^3$ variant, is metastable, but exhibits exceptional stability nevertheless, because of high energy barriers to any structural and chemical changes between the competing phases. Diamond exhibits an exceptional thermal conductivity, and p-type doping and hole conductivity has been demonstrated in the bulk, and more readily on the crystal surface.

Pure binary BN has not been investigated in as much detail as diamond, graphene, or graphite, but potentially hosts remarkable undiscovered electronic, photonic, and thermal properties. The lightest atom materials (diamond, BN) by virtue of extremely strong chemical bonds, are refractory in nature, demonstrating exceptional stability to chemical and physical perturbations such as high temperature, corrosive chemical environments, and mechanical deformation. For the same reasons, they are difficult to synthesize and control, requiring extreme environments for high quality crystal growth.

The 2D layered form, hexagonal BN (hBN) has been investigated more than the cubic phase of the crystal. Around the turn of the century in 2001, growth of single crystal cubic BN was achieved under high pressure/high temperature (HPHT) conditions by Taniguchi et al. using Li and Ba-solvent methods[14]. Soon thereafter, single crystals of the hexagonal form of boron nitride (hBN) were obtained by the Ba-B-N HPHT growth conditions. Multilayer hBN obtained in this fashion exhibited strong cathodoluminescence at ~5.8 eV and lasing under electron beam pumping[15]. Metal-organic chemical vapor deposition (MOCVD) has been used to grow multilayer hBN on sapphire, exhibiting deep-UV photoluminescence[16]. More recently, ultra high-temperature molecular beam epitaxy (MBE) has been used for the growth of hBN on graphite at thermocouple temperatures ~1300-1700 C exhibiting strong luminescence at ~5.5 eV with phonon replicas and weak sub-bandgap defect emission[17,18]. The high-temperature MBE grown hBN exhibit properties closer to the single-crystal hBN obtained by the HPHT method. High quality hBN has also been prepared by the molten metal-flux method using Ni/Cr with isotope control of Boron exhibiting shifts of PL peaks of bandgap energy[19].

Because monolayer hBN is made of $sp^2$ chemical bonds, it can form either planar 2D sheets, or can roll up into nanotubes, in close analogy to graphene and carbon nanotubes. As opposed to the zero-bandgap, linear energy dispersion bandstructure of 2D graphene, and the sensitive dependence of the electronic properties on the radius and chirality of carbon nanotubes, hBN is an electrical insulator in both the 2D planar form and in the quasi-1D nanotube forms. Multilayer hBN has been found to exhibit an indirect energy bandgap of ~5.96 eV[20], though with unexpectedly high internal quantum efficiency compared to common indirect gap semiconductors[21].

Though synthetic methods for obtaining cubic BN were known to exist in the past[22], cBN crystals of few mm sizes were recently obtained by the catalyst-assisted HPHT growth process[14]. Cathodoluminescence spectroscopy of colorless cBN crystals show emission around 200 nm. Electrical conductivity has been reported in these single crystals that are doped with Si or S (n-type), or doped with Be (p-type). Epitaxial growth of cBN is in its infancy. Very recently, radical-assisted MBE growth on (001) and (111) Diamond single crystals has been reported[23-26]. Epitaxial growth of cBN is expected to be a next frontier for nitride materials science and is expected to yield rich rewards.

hBN has intriguing photonic and electronic properties that can offer a host of potential applications. To date, hBN layers exfoliated from the bulk have been used in the field of 2D materials with graphene, $MoS_2$, and other layered semiconductors as an insulating barrier, or atomically smooth and chemically clean substrate. The strong chemical bonds of hBN also result in long-lived phonon-polariton modes for conversion of electromagnetic waves in the optical frequencies into mechanical vibrational waves in the crystal[25]. Because of the weak van der Waals bonding between layers, it has been used as a nucleation layer on sapphire for the growth of nitride LEDs and HEMTs that could be "peeled-off" and transferred to other flexible, or functional substrates[26]. Because hBN does not have spontaneous polarization, it has been used as a new reference to re-evaluate the spontaneous and piezoelectric coefficients of the nitride semiconductors[27]. A point to note is that monolayer hBN has extremely high in-plane piezoelectricity. The optical properties of hBN can be exploited in deep-UV photonics, and if electrical conductivity can be achieved in very thin layers, it can offer a new platform for high voltage and high-speed electronics taking advantage of unique scaling properties rooted in its 2D nature.

On its own, the electronic and photonic properties of cBN promise significant surprises due to the light atomic masses and strong electron-phonon coupling, and are expected to be uncovered with advances in crystal and epitaxial growth over the next decade. In the nearer term, cBN offers a large bandgap and a high thermal conductivity comparable to diamond, to potentially take nitride RF electronics into power regimes currently limited by the highest thermal conductivity of the substrate on which the devices are realized. Diamond has been investigated and found attractive for heat removal in high power GaN HEMTs, and cBN can offer a potential *epitaxial* solution to this problem.

## 3. Nitride Extreme-Piezoelectrics and Ferroelectrics

Crystalline AlN by itself is an attractive piezoelectric material capable of sustaining high temperatures, with a piezoelectric coefficient $e_{33} \sim 146 \mu C/cm^2$ , and boasting a spontaneous polarization of $P_{sp} \sim 8.1 \mu C/cm^2$. The corresponding calculated values for GaN is $e_{33} \sim 73 \mu C/cm^2$ and $P_{sp} \sim 2.9 \mu C/cm^2$ and for InN is $e_{33} \sim 97 \mu C/cm^2$ and $P_{sp} \sim 3.2 \mu C/cm^2$.[7] These values were calculated by the application of the (then-new) Berry phase formalism of electronic polarization, by comparison to a zinc-blende crystal as the reference. In most experiments, the *differences* in polarization across heterojunctions manifest as internal electric fields or mobile electron or hole gases at heterojunctions, and are in acceptable agreement with the calculated values. A new calculation that uses the layered (hexagonal) lattice as the non-polar reference has revised these coefficients significantly; in the new values, the differences in polarization are similar to the old values, but the *absolute* values are significantly different.[27]

In addition to the effects of the polarization fields in electronic and photonic devices, AlN today forms the basis for bulk acoustic wave (BAW) filters for high frequency RF signals, several of which go into each cellphone. The center frequency for such filters is given by $f_0 = v_s/2L$, where $v_s$ is the sound velocity in the material, and $L$ is the length of the acoustic cavity. As the frequency of communication signals increases to provide higher bandwidth data transfer, there is an increasing need for materials with piezoelectric coefficients higher than AlN. In 2002, it was predicted from first principles calculations by Takeuchi that though the transition metal containing nitride, Scandium Nitride (ScN), has a stable rocksalt structure, Scandium-containing III-V nitrides could potentially be stabilized in the wurtzite structure upon alloying with Al (or In, Ga), and thereby be piezoelectric[28]. In 2009, Akiyama et al. reported the first experimental observation of a nearly 500% increase in the piezoelectric coefficient $d_{33}$ of wurtzite $Sc_{0.4}Al_{0.6}N$ over AlN[29].

The rather remarkable experimentally observed giant enhancement in piezoelectricity may qualitatively be understood from the following arguments: Sc is a group-III transition element, and can take an oxidation state of +3, similar to Al (i.e., it is *isovalent*). Therefore, replacing some Al atoms with Sc should keep the insulating nature of AlN (i.e., not dope it). The fact that the outermost electrons of Sc are in $d-$orbitals instead of the $s$ and $p-$orbitals of Al result in a major difference when it incorporates in the crystal. Because there are 5 $d-$orbitals: $d_{xy}, d_{yz}, d_{zx}, d_{z^2} and d_{x^2-y^2}$, the crystal-field splits the energy levels of the Sc atom $d-$orbitals into $t_{2g}$ and $e_{2g}$ orbitals. As a result, the stable structure of ScN is cubic rock-salt, not wurtzite. ScN is a semiconductor of bandgap ~1 eV with a lattice constant of 4.505 Angstrom, and has been used as a better cubic substrate for GaN than Si. This is because its effective 111 lattice constant is $(\sqrt{2}/2) \cdot 4.505 \sim 3.186$ Angstrom, which has a ~0.1% lattice mismatch to 0001 GaN of lattice constant 3.189 Angstrom, compared to the ~17% lattice mismatch with 111 Silicon (of lattice constant 3.84 Angstrom). Furthermore, highly conductive degenerately doped n-type and p-type semiconducting ScN have also been realized[30,31]. Because of the cubic crystal structure of ScN, one would expect that the effect of incorporating Sc into AlN should be to a) distort the lattice structure from wurtzite towards cubic, and b) shrink its bandgap. Furthermore, the chemical bonding becomes increasingly *soft*: this implies a more compliant crystal that undergoes far more atomic or lattice displacements for the same strain, boosting its piezoelectricity.

Since the initial report[20] of Akiyama et al., there has been substantial interest in increasing the piezoelectric polarization of AlN by alloying transition metals. The group III transition metal below Sc in the periodic table, Yttrium (Y) has been suggested from first principles to lead to a high electromechanical coupling coefficient when substituting Al to form YAlN. Alloying in small amounts of B, which shares the +3 oxidation state, but forms much stronger chemical bonds with N, is predicted to counteract the decrease in stiffness caused in the crystal by Sc or Y substitution alone. This is expected to maintain a high electromechanical coupling coefficient, which is attractive for BAW filters[32]. Instead of relying on the +3 oxidation state Sc or Y isovalent with Al, it has been predicted that co-doping AlN with Mg of oxidation state +2 and Hf or Zr of oxidation state +4 in 1:1 ratio can lead to a giant enhancement in the piezoelectricity of AlN[33]. Rather remarkably, it has been experimentally found that co doping AlN with Mg and Nb leads to a 400% giant piezoelectricity enhancement[34]. This is counterintuitive, because Nb is a group 5 transition metal, and a Mg:Nb atomic ratio of 2:1 would be naively needed to maintain a net +3 oxidation state when substituting for Al sites. However, it is observed that a Mg:Nb ratio of $\sim$Mg$_{0.40}$Nb$_{0.24}$Al$_{0.36}$N is able to remain electrically insulating, and lead to the giant piezoelectric enhancement. By probing the oxidation states of Nb by XPS it was concluded that Nb atoms exist in the crystal in various oxidation states: +3, +4, and +5, pointing towards a much richer chemical bonding status than was previously believed.

These alloys of transition metals with Al(Ga,In)N should be considered as extreme-piezoelectrics (*x-piezos*), a valuable addition to the nitride family of semiconductors, with potentially far-reaching consequences. Because the spontaneous and piezoelectric polarization discontinuity in nitride semiconductor heterostructures are the basis for RF and power electronics, boosting the piezoelectricity by $\sim$500% can lead to several novel device applications that combine the high current carrying capacity with high breakdown voltages. The first steps to integrate ScAlN in HEMTs has been taken by MBE growth. HEMTs using such ScAlN barriers have been realized recently[35], and promise an exciting foray into unknown territories for semiconductor device electronics.

A natural question to ask is: how high can one ramp up the piezoelectricity? With the substitution of Al by Sc in AlN, the piezoelectric constant increases, with a concomitant softening of the crystal. AlN can be either Al-polar, or N-polar, it has two stable crystal structures, something that an external electric field cannot change due to very stiff bonds and a high energy barrier. Replacing Al with increasing amounts of Sc moves it closer to a cubic structure, while softening the bonds. With the addition of sufficiently large amount of Sc, can one obtain a crystal structure in which the ScAlN is so close to being cubic with softened bonding, that an external electric field can flip the structure from Al-polar to N-polar? If this happens, ScAlN can become a *ferroelectric*! Ferroelectrics, one of the newest material families (discovered in 1920), have typically been limited to oxide crystals, several of which are of the perovskite lattice structure. There were no known nitride ferroelectrics to date; the only one predicted from a high-throughput first principles calculations[w] of a vast number of possible nitride compounds is the perovskite nitride LaWN, - a material that is awaiting synthesis and characterization. As amazing as it may sound, it seems we have the glimmers of a major breakthrough in ScAlN, with extremely high piezoelectricity and potentially also ferroelectricity.

Every ferroelectric is a piezoelectric, but not all piezoelectric are ferroelectrics. For example, AlN is a piezoelectric, but not a ferroelectric. That Sc containing nitride semiconductors can potentially exhibit ferroelectricity was anticipated in first principle calculations[w]. An experimental observation of ferroelectricity has very recently been reported in ScAlN[y]. Not only is the ferroelectricity observed in ScAlN, it seems to be *very strong*, at ~100 $\mu C/cm^2$ (much higher than PbZrTiO, or PZT), and boasts coercive fields of ~1 MV/cm, and breakdown fields in excess of a few MV/cm, something that is currently unachievable in established ferroelectrics.

If stable, and repeatable ferroelectricity can be tamed and brought to the nitride platform, it opens up vast areas of applications. Ferroelectric *memory* can significantly add to the value chain created by nitride electronics. Like magnets, ferroelectrics can potentially provide significant opportunities for energy storage in power electronics. Layers of ferroelectrics can reduce short channel effects in ultrascaled transistors by an effective negative capacitance at the drain side. Energy-efficient sub-Boltzmann transistors that require less than 60 mV/decade can potentially be realized using negative capacitance in the gate of FETs. Since ferroelectric layers can be poled by an electric field, reconfigurable periodically poled arrays can be used for second harmonic generation and new sorts of devices in non-linear optics. When combined with the electronic and optical nitride devices, the possibilities enabled by epitaxial ferroelectric layers seamlessly integrated with transistors, LEDs, and Lasers are limited only by our imagination.

## 4. Magnetic Nitrides

Nitride RF and power electronics platforms can benefit tremendously if magnetic materials could be integrated seamlessly with the GaN material system. For example, in power electronics, the storage of electrical energy occurs in magnets and inductors, and RF systems require circulators and non-reciprocal elements that are magnetic (e.g. magnetic insulators, or *ferrites*). Magnetism is a phenomena originating in the *spin* of fermionic particles such as electrons. Spin is the intrinsic angular momentum of all fermions that has no classical analog. The nucleus hosts neutrons and protons that are fermions, each of spin $1/2$, or spin angular momentum $\hbar/2$. Each electron in the core energy levels as well as those participating in the formation of bonds and bands of a crystal possess the same spin angular momentum of $\hbar/2$. Adding up the total nuclear and electronic spin angular momentum vectors in crystals of typical group IV semiconductors, III-V semiconductors GaAs, GaN (or AlN and InN), or II-VI semiconductors such as CdTe, results in a net zero spin: meaning they are non-magnetic materials (actually they are very weakly diamagnetic, meaning they oppose an external magnetic field). In 1996, it was discovered by Ohno et al. that adding dilute concentrations of the transition metal Mn to GaAs made the alloy GaMnAs *ferromagnetic* at low temperatures[39]. The existence of Mn-doped dilute magnetic semiconductors was developed in II-VI semiconductors such as CdMnTe well before the discovery of GaMnAs[40], but they are typically not ferromagnetic: they exhibit paramagnetic, antiferromagnetic, or spin-glass behavior. A crucial difference between Mn-based II-VI and III-V dilute magnetic semiconductors is in their *electronic* conductivity: for example, CdMnTe is electrically insulating, whereas GaMnAs has p-type conductivity. The holes in GaMnAs are responsible for carrier-mediated ferromagnetism, which is absent in CdMnTe due to its lack of mobile carriers.

The transition metal Mn has the electronic orbital structure $[Ar]3d^5 4s^2$, with 5 electrons in the $d$-shell: it is half-filled. In a II-VI semiconductor such as $Cd_{1-x}Mn_xTe$, when Mn substitutes a group II metal site, the two $4s$-orbital electrons of Mn participate in chemical bonding, leaving the 5 $d$-orbital electrons highly localized on the Mn site. These five electrons are unpaired with spin configuration ↑↑↑↑↑, leading to a net spin angular momentum of $S = 5\hbar/2$ per Mn atom. Mn thus incorporates in the +2 oxidation state, and is *isovalent* with the group II metal it replaces (e.g. Cd) in II-VI semiconductors, and does not introduce extra mobile carriers, it is not a dopant. Since the spins on Mn sites are localized and do not have strong interactions with neighboring Mn sites, the $Cd_{1-x}Mn_xTe$ crystal is not able to become ferromagnetic. On the other hand, in GaMnAs, Mn also exists in the +2 oxidation state, and thus acts as an *acceptor* dopant. Since Mn introduces an unoccupied energy level close to the valence band edge of GaAs, it introduces holes in the valence band, making it p-type. The mobile holes mediate the interactions between Mn spin centers, making the entire GaMnAs crystal ferromagnetic. The critical (Curie) temperature for such carrier-mediated ferromagnetism is given by $T_c \approx C N_{Mn} \beta^2 m_v^* p^{1/3}$, where $C$ is a constant, $N_{Mn}$ is the concentration of Mn spins, $\beta$ is an exchange interaction constant (called the $p - d$ coupling between mobile holes in $p$ −states and localized spin $d$ −states at Mn sites), $m_v^*$ is the effective mass of holes, and $p$ is the hole concentration. To date, a maximum $T_c \sim 200$ K has been achieved in GaMnAs, which falls short of room temperature for practical applications.

The hopes of achieving a room temperature dilute magnetic semiconductor in GaN by magnetic doping with Mn was high. This expectation was based on theoretical predictions based on a carrier-mediated (Ruderman-Kittel-Kasuya-Yosida) RKKY-type interactions, which is similar to a Zener model in which the Friedel-like oscillations of a typical RKKY interaction are washed out because of the large distances between the spin centers at the Mn ions, when mediated by free carriers[41]. Carrier mediated ferromagnetism in dilute magnetic semiconductors such as in GaMnAs occurs in conductive material with free carriers that get spin-polarized due to their interaction with the magnetic sublattice of the Mn sites. However, it is believed that Mn in the +2 oxidation state introduces an energy level that is rather deep in the gap of GaN, and as a result is unable to provide mobile holes: GaMnN is found to be semi-insulating. The Curie temperature for ferromagnetism is experimentally observed to be rather low, about $T_c \sim 1$ K at a Mn concentration of ∼3%. The mechanism is thought to be by short-range ferromagnetic superexchange interactions. Compared to GaMnAs, the GaN crystal possesses a far higher piezoelectric effect. It is then possible to use the piezoelectricity to control the magnetic interactions, as was shown recently[42]. This is an indirect electrical control of magnetism, mediated by the strain and piezomagnetism. Even though the interaction is weak and occurs at low temperatures, this is a *magnetoelectric* property, because the magnetization can be controlled, via the strain by electric fields. It would be of high interest if stronger magnetoelectric properties closer to room temperature could be found in nitride materials in the future.

Instead of alloying Mn with GaN, it is possible to create binary phases of its nitride: for example, MnN, and $Mn_xN$. MnN (called the $\theta$-phase), has a cubic crystal structure, in which Mn is in a +3 oxidation state and N in the -3 state, similar to a III-V crystal. MnN crystals are antiferromagnetic. $Mn_xN$ on the other hand (the $\epsilon$-phase) has an fcc anti-perovskite crystal structure, and is found to be electrically conducting (metallic), and *ferrimagnetic*. N occupies the body-center sites, and Mn the corner sites, and the face-center sites. The Mn atoms at the corner sites have magnetic moments of $3.85\mu_B$ per formula unit, and the face-center Mn atoms have a magnetic moment of $-0.9\mu_B$ per formula unit aligned along the 100 (c-) axis. The ferrimagnetic transition temperature of $Mn_xN$ is $T_c \sim 740\,K$, well above room temperature[43]. When deposited on nearly lattice-matched $SrTiO_3$ substrates, $Mn_xN$ exhibits remarkably large magnetic domains, approaching millimeters in size[44]. The magnetization points perpendicular to the plane of the thin film, which makes it attractive for high-density magnetic storage, with spin-torque switching for non-volatile energy-efficient memory applications. It may be considered for power electronic circuits in the future. Finally, the rare earth nitride crystal GdN is ferromagnetic with a Curie temperature $T_c = 70\,K$, and EuN has $T_c = 120\,K$. Unlike Mn containing Nitrides in which the magnetism originates from $d$-shell electrons, in GdN and EuN the magnetism is due to unpaired spins of electrons in the $f$-shell[45-47].

## 5. Metallic and Superconducting Nitrides

The transition metals in the groups headed by Ti, V, and Cr in the periodic table all form highly refractory and chemically stable nitride *compound* metals. Several of them are heavily used in electronics (e.g. WN for contacts in Silicon technology). The room temperature electrical resistivities of typical transition metal nitrides range from $\rho \sim 7\mu\Omega \cdot cm$ for TiN and ZrN to $\rho \sim 200\mu\Omega \cdot cm$ for TaN and CrN[48,49]. For comparison, the room temperature electrical resistivity of the highest conductivity metals Cu and Ag is $\rho \sim 1.5\mu\Omega \cdot cm$. When the compound metallic nitrides TiN, ZrN, HfN, VN, NbN, TaN, MoN, and WN are cooled to cryogenic temperatures, they undergo an electronic phase transition from the metallic to the superconducting phase. Among them, NbN shows the highest superconducting transition temperature at $T_c \sim 17\,K$. These properties of metallic and superconducting transition metal nitrides have been investigated heavily in the past few decades. The chemical bonding is between the $d$-orbitals of the transition metal elements, and the outermost $p$-orbitals of N. Most of these crystals have a stable rocksalt crystal structure, though hexagonal forms can also exist. The normal metallic state resistivities, and the superconducting transition temperatures are sensitive to the stoichiometry, impurity densities, and the lattice constants of these nitride crystals.

Most of these metallic and superconducting nitrides have been grown in the thin film form in the past by RF magnetron sputtering based techniques, or more recently by atomic layer deposition (ALD). The III-Nitride semiconductor material heterostructures are grown by epitaxial techniques with electronic grade purity and control. The development and high level of maturity of Metal-Organic Chemical Vapor Deposition (MOCVD) and Molecular Beam Epitaxy (MBE) of nitride materials today offers a tantalizing possibility of re-examining the growth of transition metal nitride crystals at a level of purity, thickness, and compositional control commensurate with electronic grade semiconductors. In addition, an exciting possibility is their potential *epitaxial* integration with the III-nitride semiconductors to form metal/semiconductor and superconductor/semiconductor heterostructures.

The closeness of the lattice constant of metallic cubic TiN and semiconducting cubic AlScN has recently enabled the epitaxial growth of low defect-density metal/semiconductor nitride multilayers[49]. In a similar vein, the (111) lattice constant of the cubic crystal structure of NbN, HfN, and other transition metal nitrides is close to the $c$-plane $a$-lattice constants of wurtzite SiC, AlN, and GaN, and offers the required hexagonal symmetry. This has led to recent demonstrations of epitaxial growth of heterostructures of NbN, SiC, AlN, and GaN by MBE[50]. The transition metal nitride layers offer several features that can be exploited in future generations of devices. Similar to layered hBN nucleation layers, they offer a way to be selectively etched chemically, allowing complete liftoff of epilayers or processed devices on top[51]. Epitaxial transition metal nitride epitaxial layers of desired thicknesses down to a few nanometers can be inserted in GaN/AlGaN/InGaN heterostructures for various electronic, photonic, and plasmonic device applications. For example, recently it was shown that AlGaN/GaN layers epitaxially grown on thin NbN layers exhibited high mobility 2DEGs exhibiting quantum oscillations in magnetotransport, and could be processed into high performance HEMTs interfaced with the superconducting properties of NbN[52]. This leads to the feasibility of epitaxial metal/semiconductor Schottky diodes and epitaxial gate junctions for transistors. Buried epitaxial metal (epimetal) layers of low resistivity can be inserted if needed for electric field management in high-voltage or high-speed electronic devices. Low resistivity epimetals could potentially be used for mirrors and reflective contacts, and for plasmonic applications in conjunction with GaN photonic devices.

At low temperatures, the metal→superconductor transition of nitride layers offer several unique possibilities and device applications. Current transport in the superconducting regime in for example NbN is a genuine dissipation-less mechanism, when a finite current flows through the superconducting layer for zero voltage drop. The physical mechanism responsible for standard superconductors (such as the transition metal nitride superconductors) is the pairing of electrons to form Cooper pairs, and their condensation to a correlated macroscopic quantum state below the critical temperature. The critical magnetic fields that NbN can sustain before losing superconductivity can be in excess of 15-20 T - which makes them very robust against phase transition due to the Meissner-Ochsenfeld effect. A finite energy is required to break the Cooper-pairing, characterized by an energy $\Delta \sim k_b T_c$ , where $T_c$ is the critical temperature; this may be thought of similar to the bandgap of a semiconductor, and is indeed referred to as the superconducting *gap*. For example, when a photon of energy larger than the superconducting gap is incident on the superconductor, it causes a localized superconductor→metal transition (similar to an electron-hole pair generation by a photon in semiconductor).

In 2001, it was discovered that narrow NbN superconducting strips could use the above mechanism for ultrafast detection of *single photons*[43]. Because a finite voltage must drop across a metal to carry a current, the event of the detection of a single photon appears as a voltage pulse that dies out within picoseconds as the superconductor recovers. This mechanism is now used for NbN based single-photon detection, counting, and imaging at high data rates and high efficiencies, as long as the detector is cooled sufficiently below $T_c$. Such nitride superconducting detectors are expected to play an increasingly important role in the future for single-photon secure quantum communications, and also for astronomy. Epitaxial integration of such superconducting NbN layers with active nitride electronic and photonic emitter structures (e.g. single photon emitters) can lead to new forms of single photon detection paradigms, and to a unique platform for secure quantum communications.

Epitaxial nitride metals can potentially serve as buried interconnect layers in structures that have several transistor layers vertically stacked for 3D integration. These interconnects can become lossless at temperatures below which the nitride metals become superconducting. The possibility of epitaxial growth of superconductor/insulator/superconductor tunnel junctions (e.g. NbN/AlN/NbN) offers a possibility to realize Josephson junctions with low variation of properties that depend sensitively on the tunnel barrier thickness, and its chemical, electronic, and structural properties. Such Josephson junctions have been grown epitaxially in the past by sputtering, proving their feasibility[34]. Epitaxial growth offers the possibility of scalability to large wafer sizes, and the potential integration with GaN electronic devices such as RF amplifiers, transistor switches, AlN based micro-electro-mechanical systems (MEMs) and acoustic wave structures. These building blocks offer opportunities for superconducting digital (von-Neumann) electronics using established paradigms such as the Rapid Single-Flux Quantum (RFSQ)[35] significantly enhanced toolset than in the past. More interestingly, the combination of nitride Josephson junctions with NbN based transmission lines for superconducting qubits, NbN single photon detectors, GaN HEMT based RF amplifiers, and AlN based acoustic wave structures provide the ingredients necessary for a all-nitride platform for quantum computation and communications[36-37].

The combination of the high RF powers generated by GaN HEMTs today, with the potential for superconducting NbN based lossless microwave cavities offer the potential for superconducting RF (SRF) cavities that generate high powers in very narrow frequency ranges with quality factors in the billions. Such devices are of interest in on-chip precision clocks, particle accelerators, and also for various medical applications. Of course, for all such applications the nitride superconducting materials must be cooled below the critical temperature, the highest of which is $T_c \sim 17\,K$ for NbN. We are still far from higher temperature nitride superconductors. In 1987, a new class of "high-$T_c$" superconductors were discovered in *layered* oxides, called the cuprates (such as Yttrium Barium Copper Oxides)[38]. The highest critical temperatures such "high-$T_c$" superconductors today stands at ~150K. Though this is still far below room temperature, it exceeds the boiling point of liquid nitrogen of 77K. These superconductors are composed of conducting 2D $CuO_2$ layers separated by interlayers. The interlayers modulation-dope the $CuO_2$ layers with electron or hole carriers. Can we have "high-$T_c$" nitride superconductors, similar to the cuprates? Ten years after the discovery of the cuprates, it was discovered that *layered* HfN combined with halide interlayers, and modulation doping from Lithium[39] resulted in doped HfNCl:Li structures that exhibited a superconducting transition temperature of $T_c$~25.5 K, higher than the standard nitride metals. This is not too high yet, but it follows the trajectory of the cuprates. So there is reason to be optimistic for designer higher-$T_c$ nitride superconductors in the future.

## 6. Conclusions

To conclude, in this article we have identified and reviewed four families of emerging nitride materials that have the potential to expand the capabilities of the currently mature GaN, InN, and AlN semiconductors. In these families indicated in Figs. 2 and 3 are BN and ScN as emerging *semiconductors*, Sc- and Y-based nitrides are highly *piezoelectric* (and potentially *ferroelectric*) materials, Mn and rare-earth based nitrides are *magnetic*, and transition metal nitrides are excellent *metals* and *superconductors*. It is possible to *epitaxially* integrate these disparate material families to create low defect and high quality heterostructures of semiconductors, ferroelectrics, magnets, and superconductors. Because of the richness of these material properties, the new nitrides have the potential to significantly expand the electronic and photonic applications space of nitride materials in the classical information systems for computation, memory, and communications. Furthermore, because of the confluence of these material properties in a single epitaxial nitride material platform, it offers unique opportunities for quantum information systems of the future.

## Acknowledgments

The authors acknowledge financial support from the Semiconductor Research Corporation (SRC/DARPA) Joint University Microelectronics Program (JUMP), an Office of Naval Research grant (#N00014-17-1-2414) monitored by Dr. Paul Maki, a US Air Force Office of Scientific Research grant (AFOSR FA9550-17-1-0048) monitored by Dr. Ken Goretta, a NSF EFRI-2DARE (#1433490), a NSF DMREF (#1534303), a NSF E2CDA (#1740286), and a NSF RAISE TAQs (#1839196) monitored by Dr. D. Dagenais. The authors are grateful for a research collaboration with Dr. David Meyer and his research group at the Naval Research Laboratory. Discussions with Profs. Darrell Schlom, Dan Ralph, Bruce Van Dover, Bob Buhrman, Farhan Rana, David Muller, Katja Nowack, and Greg Fuchs of Cornell University, Prof. Guillaume Cassabois of the University of Montpellier, Prof. Sergei Novikov of the University of Nottingham, Prof. Hong Tang of Yale University, Prof. Dr. Detlef Hommel and Dr. Edyta Piskorska of the Polish Academy of Sciences, Dr. Takashi Taniguchi of NIMS (Japan), Dr. Tom Kazior of Raytheon, Dr. Han Wui Then of Intel, Dr. Young-Kai Chen of DARPA, and Dr. Yu Cao of Qorvo are gratefully acknowledged.

## Figure Captions

**Fig. 1.** The elements contained in traditional nitride semiconductor technology are shown with those in the "New" Nitrides in the periodic table. The color coding indicates the families of physical properties or usage of the elements in nitride crystals.

**Fig. 2.** Schematic figures showing the new nitride materials such as Boron-containing hBN and cBN, Sc-containing nitrides, and transition metal nitrides. The piezoelectric coefficients of several materials are shown against the maximum processing temperatures, adapted from[29].

**Fig. 3.** A summary of the four broad categories of nitride materials. The currently known candidates for semiconducting properties, metallic and superconducting properties, magnetic properties, and high piezoelectric and ferroelectric properties are listed in the figure. Epitaxial integration offers the potential to integrate across families, such as the semiconductor/superconductor heterostructures of GaN/NbN, or the metal/insulator or metal/x-piezo heterostructures of TiN/ScAlN.

Fig.1.

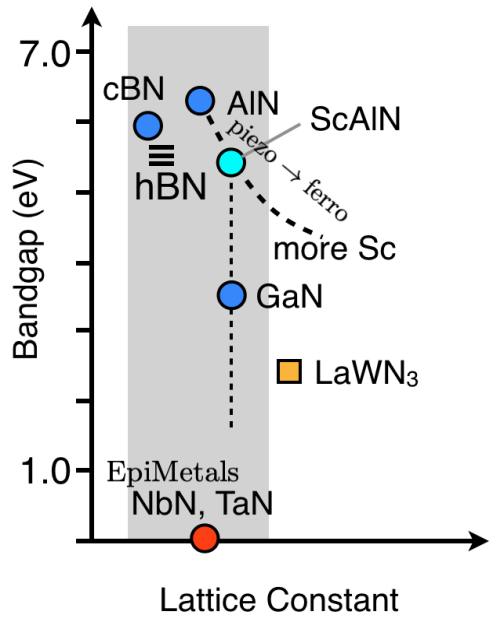 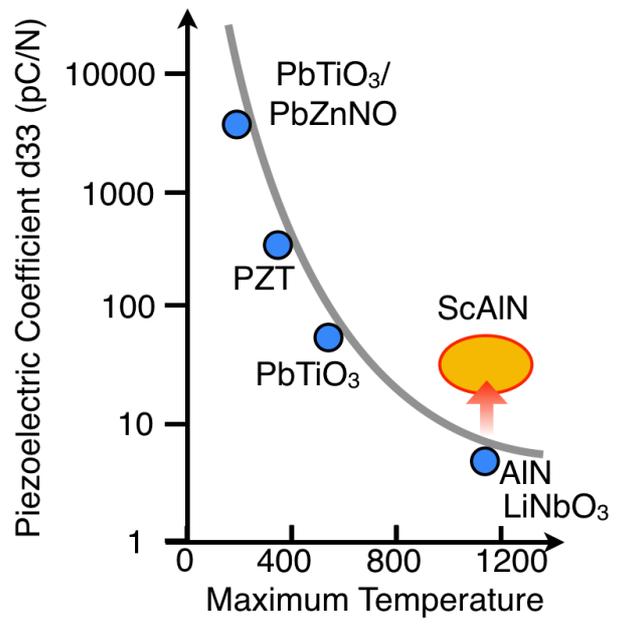

Fig. 2.

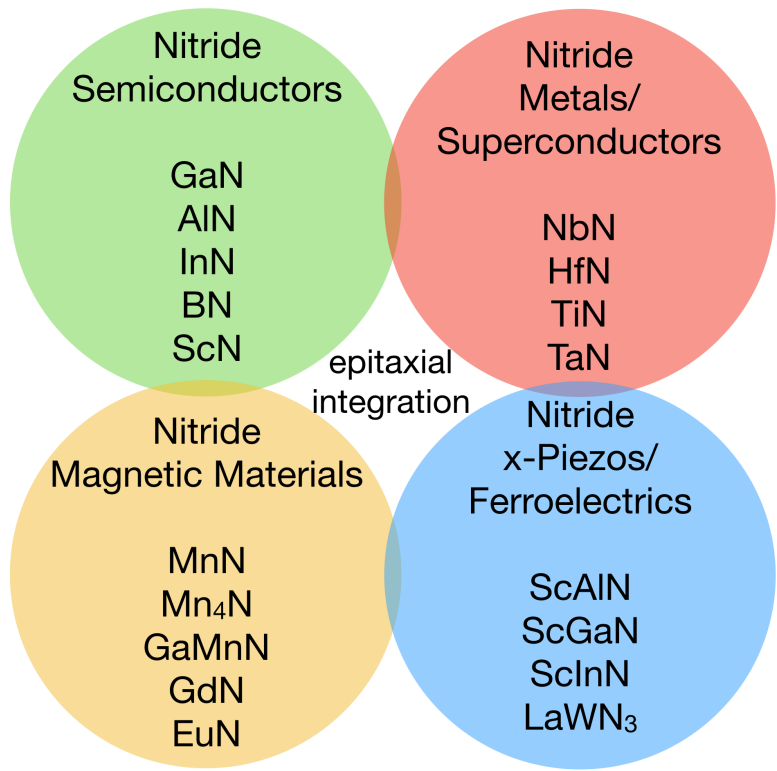

Fig. 3.